\documentclass[aps,prl,reprint,superscriptaddress,nofootinbib,floatfix]{revtex4-2}

\usepackage{amsmath,amssymb,bm}
\usepackage{graphicx}

\usepackage{xcolor}
\definecolor{revisionblue}{RGB}{0,70,155}
\newif\ifshowrevisions
\ifdefined\BellCleanCopy
  \showrevisionsfalse
\else
  \showrevisionstrue
\fi

\newcommand{\lambdabarc}{\bar{\lambda}_c}

\begin{document}
\title{Separation-Dependent Bell Correlations from Prepared Coherence in Dirac Wavepackets}

\author{Ju Gao}
\email{jugao@illinois.edu}
\author{Fang Shen}
\affiliation{Department of Electrical and Computer Engineering, University of Illinois at Urbana–Champaign, Urbana, Illinois 61801, USA}

{\date{October 29, 2026}

% V3 C01: Introduce physical wave entities with the packets; state interference and local sampling without naming LWR.
\begin{abstract}
We calculate separation-dependent Bell correlations for a fixed pair
of Dirac electron wavepackets in a free, weakly relativistic packet
model, treating the prepared joint wave as a physical wave--entity.
Each station separation is sampled at the corresponding central packet
arrival time. For the illustrated phase \(\chi=\pi/4\), the magnitude
of the prescribed-setting Bell--CHSH response decreases and revives;
optimizing the spin measurement directions retains the modulation.
It occurs already at leading spin order, including for equal prepared
branch weights. Interference between direct and exchanged assignments
of the electron waves to the stations changes the relative weights
and phases of the sampled spin amplitudes, thereby modulating the
optimized response. Representative finite spatial and timing windows
reduce visibility while retaining broader suppression and revival.
This wave--entity account gives a quantitative, testable correlation
pattern from prepared coherence, deterministic propagation, and local
sampling, without postulating transmission between detectors.
\end{abstract}

\maketitle

% V3 C02: Focus motivation and relevant relativistic literature; broad technology references are assessed in response Point 4.
\section{Introduction}

Bell--CHSH correlations are usually formulated for an ideal spin
singlet, whose optimized response reaches \(2\sqrt2\)
\cite{Bell1964,CHSH1969,Tsirelson1980}. CHSH violation excludes
Bell-local outcome factorization under the usual independence and
sampling assumptions, as tested in loophole-free experiments
\cite{Brunner2014,Hensen2015,Giustina2015,Shalm2015}.
Here we retain the spatial structure of a prepared Dirac pair to ask
how free propagation and local sampling determine the Bell response
as the measurement stations are moved apart.

Relativistic quantum-information studies show that momentum spread,
Lorentz transformations, and the choice of spin observable affect
reduced-spin entanglement and EPR--Bohm correlations
\cite{Czachor1997,Peres2002,GingrichAdami2002,CabanRembielinski2005,
FoldyWouthuysen1950,Bauke2014Spin}. Bell optimization for massive
particles \cite{Streiter2021}, localized detector models
\cite{Perche2024}, and localization-conditioned Dirac-pair correlations
\cite{Caban2014Localization} further motivate retaining the measurement
geometry. Recent work also examines Lorentz-boost effects on Gaussian
wavepacket coherence \cite{Mukherjee2024Coherence} and on discord and
entropic correlations of two relativistic fermions
\cite{Kurashvili2022Discord}.\label{loc:rqi_v2}
Our question concerns a fixed preparation, coherent direct and exchanged
endpoint assignments, and their separation-dependent CHSH response
under prescribed and optimized spin measurements.

In the wave--entity (WE) description
\cite{EntropyEvaSpin2024,GaoShen2025_JPhysComm}, the electron is a spatially
extended physical Dirac wave. The prepared pair is represented by
an antisymmetrized, generally nonseparable joint wave in an
opposite-spin sector. In the ideal spin-only description with
orthogonal propagation modes, generalizing the Bohm--EPR presentation
\cite{BohmAharonov1957}, the source state is
\begin{equation}
|\Psi_0\rangle
=
\cos\theta\,|\uparrow_A\downarrow_B\rangle
-
e^{i\chi}\sin\theta\,|\downarrow_A\uparrow_B\rangle ,
\end{equation}
where \(\theta\) and \(\chi\) fix the relative branch weight and phase.
The labels \(A,B\) denote propagation modes associated with the intended
stations, not distinguishable electron identities. In this spin-only
description, \(\theta=\pi/4\) gives maximal entanglement and
\(0<\theta<\pi/4\) partial entanglement, with preparation concurrence
\(C_{\rm prep}=|\sin(2\theta)|\). We call these fixed preparations
``balanced'' and ``imbalanced''; their concurrence does not determine
the spin state conditioned on later endpoint sampling. The complete
spatial state is antisymmetrized before propagation and sampling.

Propagation to the planes at \(\pm Z\), followed by fixed,
setting-independent local sampling, gives the plane-conditioned spinor
\(\mu_Z\) and the correlation chain
\[
\mu_Z\longrightarrow\boldsymbol{\tau}(Z)
\longrightarrow\{B_{\rm fixed}(Z),B_{\max}(Z)\}.
\]
Both responses use the same density-conditioned
\(\boldsymbol\Sigma\) tensor; the second optimizes the analyzer
directions. For the illustrated phase \(\chi=\pi/4\), suppression and
revival persist after optimization, including for balanced preparation.
We derive the direct--exchange mechanism, compare the two responses,
and quantify its visibility under representative finite spatial and
timing acceptance.

\section{Prepared two-Dirac-wave state}

We use a weakly relativistic, positive-energy
Dirac wavepacket approximation \cite{Dirac1928,Thaller1992} constructed
in the momentum regime
\[
\frac{\hbar}{d}\ll P\ll mc,
\]
where \(d\) is the initial packet width and \(P\) is the central
longitudinal momentum. The first inequality makes the central momentum large compared with the intrinsic packet momentum spread, while the second controls the weakly relativistic expansion. Positive energy is selected to the order retained in the spinor construction. Mutual interactions are also omitted in this analysis. 

We choose the \(z\)-axis as both the mean propagation axis and the
spin-quantization axis.  The two propagation branches have central
momenta \(+P\hat{\mathbf z}\) and \(-P\hat{\mathbf z}\), denoted by
\(R\) and \(L\), respectively.  In this basis the spin labels
\(\uparrow\) and \(\downarrow\) refer to the prepared spin polarization
along the \(z\)-axis.  For strictly longitudinal positive-energy Dirac
plane waves, this spin label is preserved under free propagation.  For
finite Gaussian packets, the transverse momentum spread also enters the
lower Dirac components displayed below. The branch spin labels refer
to the dominant upper-component polarizations.

A single right-moving Dirac wavepacket is written as

{\small\[
\psi_{\uparrow}(P;\mathbf r,t)
=
A(\mathbf r,t)R_{\uparrow}(\mathbf r,t),
\,\,
\psi_{\downarrow}(P;\mathbf r,t)
=
A(\mathbf r,t)R_{\downarrow}(\mathbf r,t),
\]
}

and the corresponding left-moving packet as

{\small
\[
\psi_{\uparrow}(-P;\mathbf r,t)
=
A(\mathbf r,t)L_{\uparrow}(\mathbf r,t),
\,\,
\psi_{\downarrow}(-P;\mathbf r,t)
=
A(\mathbf r,t)L_{\downarrow}(\mathbf r,t).
\]
}

Introduce the dispersion scale, complex Gaussian width, and
longitudinal phase-gradient length
\begin{equation}
L^2(t)=\frac{\hbar t}{m},
\quad
D^2(t)=d^2+iL^2(t),
\quad
\mathcal L_P=\frac{d^2P}{\hbar}.
\label{eq:packet_scales_general}
\end{equation}
The common scalar envelope is
\begin{eqnarray}
A(\mathbf r,t)
&=&
\left[
\frac{d}{\sqrt{\pi}D^2(t)}
\right]^{3/2}
e^{-imc^2t/\hbar}
\nonumber\\
&\times &
\exp\left[
 -\frac{r^2+iL^2(t)\mathcal L_P^2/d^2}{2D^2(t)}
\right].
\end{eqnarray}

We further denote the reduced Compton wavelength by
\[
\lambdabarc\equiv\frac{\hbar}{mc}.
\]
The spinor factors contain a common branch phase and the leading
positive-energy lower components.  Writing
\[
\begin{aligned}
R_s(\mathbf r,t)
&=\mathcal R_s(\mathbf r,t)
\exp\left[\frac{i\mathcal L_Pz}{D^2(t)}\right],\\
L_s(\mathbf r,t)
&=\mathcal L_s(\mathbf r,t)
\exp\left[\frac{-i\mathcal L_Pz}{D^2(t)}\right],
\end{aligned}
\]
for \(s=\uparrow,\downarrow\), we use
\[
\mathcal R_{\uparrow}(\mathbf r,t)
=
\begin{pmatrix}
1\\
0\\
\dfrac{\lambdabarc}{2}
\dfrac{iz+\mathcal L_P}{D^2(t)}\\
\dfrac{\lambdabarc}{2}
\dfrac{ix-y}{D^2(t)}
\end{pmatrix},
\]
\[
\mathcal R_{\downarrow}(\mathbf r,t)
=
\begin{pmatrix}
0\\
1\\
\dfrac{\lambdabarc}{2}
\dfrac{ix+y}{D^2(t)}\\
\dfrac{\lambdabarc}{2}
\dfrac{-iz-\mathcal L_P}{D^2(t)}
\end{pmatrix},
\]
and
\[
\mathcal L_{\uparrow}(\mathbf r,t)
=
\begin{pmatrix}
1\\
0\\
\dfrac{\lambdabarc}{2}
\dfrac{iz-\mathcal L_P}{D^2(t)}\\
\dfrac{\lambdabarc}{2}
\dfrac{ix-y}{D^2(t)}
\end{pmatrix},
\]
\[
\mathcal L_{\downarrow}(\mathbf r,t)
=
\begin{pmatrix}
0\\
1\\
\dfrac{\lambdabarc}{2}
\dfrac{ix+y}{D^2(t)}\\
\dfrac{\lambdabarc}{2}
\dfrac{-iz+\mathcal L_P}{D^2(t)}
\end{pmatrix}.
\]
These are positive-energy Dirac spinors expanded to leading order in
\(P/mc\) and the packet momentum spread, retaining both the spatial
propagation phase and lower-component spin structure.

The general antisymmetrized entangled two-wave state is then
\begin{equation}
\Psi(\mathbf r_1,\mathbf r_2,t)
=
\mathcal N
A(\mathbf r_1,t)A(\mathbf r_2,t)
\mu(\mathbf r_1,\mathbf r_2,t),
\end{equation}
with

{\small
\begin{eqnarray}
\mu &=& \cos\theta
\left[
R_{\uparrow}(\mathbf r_1,t)\otimes L_{\downarrow}(\mathbf r_2,t)
-
L_{\downarrow}(\mathbf r_1,t)\otimes R_{\uparrow}(\mathbf r_2,t)
\right] \nonumber\\
&-&
e^{i\chi}\sin\theta
\left[
R_{\downarrow}(\mathbf r_1,t)\otimes L_{\uparrow}(\mathbf r_2,t)
-
L_{\uparrow}(\mathbf r_1,t)\otimes R_{\downarrow}(\mathbf r_2,t)
\right].\nonumber\\
\end{eqnarray}
}

Here \(\mathcal N\) normalizes the joint wave; \(\mu\) contains its
spinor-branch structure and \(A(\mathbf r_j,t)\) the common scalar
envelopes. Antisymmetrization enforces the identical-fermion structure.
The preparation parameters \(\theta,\chi\) fix the opposite-spin
branch coherence, while \(d,P\) determine the propagated endpoint
assignment amplitudes.

At leading order, each term contains one spin-\(\uparrow\) and one
spin-\(\downarrow\) upper-component branch and therefore belongs to the prepared opposite-spin sector. The balanced, phase-matched choice
\(\theta=\pi/4\), \(\chi=0\) reduces to the singlet-like sector in the
spin-only limit.  For general \(\theta\) and \(\chi\), the state retains the spatial wavepackets, exchange assignments, Dirac spinor structure, and prepared coherence that enter the endpoint spinor \(\mu_Z\).

\section{Spin correlations from endpoint sampling}

The propagated joint wave determines the endpoint-conditioned spin
tensor used for both prescribed-setting and optimized CHSH responses.

The detectors sample the planes $z_1=Z$ and $z_2=-Z$, separated by
$2Z$, through the setting-independent spatial windows
\begin{equation}\label{eq:endpoint_windows}
W_A(\mathbf r_1)=\delta(z_1-Z),
\quad
W_B(\mathbf r_2)=\delta(z_2+Z),
\end{equation}

with full-plane transverse acceptance. These delta windows define
equal-time density-slice conditioning; finite detection probabilities
require finite windows. For \(Z>0\), they are narrow-window limits of
disjoint endpoint acceptances. Spatial and timing gates and
pair-identification rules are fixed independently of analyzer settings
and spin outcomes, with both outcomes retained and no setting-dependent
detection bias.

Each window acts locally; correlations are formed by subsequently
comparing the accepted records. The coordinate labels \(1,2\) denote
arguments of the antisymmetrized wave. In the ordered sector
\(z_1=Z,z_2=-Z\), the two Bell subsystems are the endpoint spin
responses conditioned on one occupation at each station, not labeled
electron identities \cite{Schliemann2001Fermions,WisemanVaccaro2003}.

We choose the central arrival time of the quadratic-dispersion packets,
\begin{equation}\label{eq:endpoint_scales}
T=\frac{mZ}{P},
\quad
L^2(T)=\frac{\hbar Z}{P},
\quad
D^2(T)=d^2+i\frac{\hbar Z}{P}.
\end{equation}

This choice places the free-packet centers at the stations. The
preparation \((d,P,\theta,\chi)\) is fixed while \(Z\) and \(T\) vary
together: the curves describe this specified space--time sampling
protocol, not an isolated distance effect at fixed time.

Thus
\begin{equation}\label{eq:D_modulus}
|D^2(T)|^2=d^4+\frac{\hbar^2Z^2}{P^2},
\end{equation}
and the transverse Gaussian width at the sampling time is
\begin{equation}\label{eq:qZ_width}
l_Z^2\equiv\frac{|D^2(T)|^2}{d^2}
=d^2+\left(\frac{\hbar Z}{Pd}\right)^2.
\end{equation}

At the selected planes, the wavefunction is
\begin{equation}\label{eq:Psi_endpoint}
\Psi_Z=\mathcal N A_1A_2\mu_Z,
\end{equation}
where $A_j=A(\mathbf r_j,T)$ and
$\mu_Z(x_1,y_1;x_2,y_2)$ is the plane-conditioned entangled spinor.
Writing $\boldsymbol\rho_j=(x_j,y_j)$, the normalized spin-correlation
tensor is
\begin{equation}\label{eq:tau_definition}
\tau_{ij}(Z)=
\frac{
\displaystyle\int d^2\rho_1d^2\rho_2\,
|A_1A_2|^2\,
\mu_Z^\dagger(\Sigma_i\otimes\Sigma_j)\mu_Z
}{
\displaystyle\int d^2\rho_1d^2\rho_2\,
|A_1A_2|^2\,\mu_Z^\dagger\mu_Z
}.
\end{equation}
For the ideal analyzer model used here, an analyzer along the unit
direction \(\hat{\mathbf n}\) measures
\(\hat{\mathbf n}\cdot\boldsymbol{\Sigma}\), with outcomes
\(\pm1\). Joint outcome probabilities follow the Born rule for
the two local measurements, conditioned on the setting-independent
endpoint windows. The tensor in Eq.~\eqref{eq:tau_definition}
gives the corresponding spin-product averages.

The tensor combines propagation with conditioning on the two
endpoint planes. To evaluate its entries, we first isolate the
relative scalar weights of the direct and exchanged endpoint
assignments. Their propagation factors are
\(\exp[2i\mathcal L_P Z/D^2(T)]\) and
\(\exp[-2i\mathcal L_P Z/D^2(T)]\), respectively, where
\(\mathcal L_P=d^2P/\hbar\). Factoring out the direct scalar factor
leaves the relative scalar assignment amplitude
\begin{equation}\label{eq:qZ_definition}
\begin{aligned}
q_Z
&\equiv\exp\!\left[-\frac{4i\mathcal L_PZ}{D^2(T)}\right]
=\rho_Ze^{-i\Theta_Z},\\
\rho_Z
&=\exp\!\left(-\frac{4Z^2}{l_Z^2}\right),
\qquad
\Theta_Z=\frac{4d^2PZ}{\hbar l_Z^2},
\end{aligned}
\end{equation}
where \(l_Z^2=|D^2(T)|^2/d^2\). The magnitude \(\rho_Z\) and phase
\(\Theta_Z\) specify the exchanged contribution relative to the direct
one at the sampled endpoints; \(q_Z\) is not a global wavepacket
overlap. Both assignments' spinor factors are retained below.

The remaining endpoint quantities enter the Dirac lower components:
\begin{equation}\label{eq:phi_eta_Zpm}
\eta_Z\equiv\frac{\lambdabarc}{2D^2(T)},
\quad
\mathcal Z_\pm\equiv Z\pm i\mathcal L_P.
\end{equation}
Collecting the prepared amplitudes and their exchanged contributions
gives
\begin{equation}\label{eq:ABXYdefinitions}
\begin{aligned}
A_Z&=\cos\theta+e^{i\chi}\sin\theta\,q_Z,\\
B_Z&=e^{i\chi}\sin\theta+\cos\theta\,q_Z,\\
X_Z&=\cos\theta\,\mathcal Z_-
     +e^{i\chi}\sin\theta\,q_Z\mathcal Z_+,\\
Y_Z&=e^{i\chi}\sin\theta\,\mathcal Z_-
     +\cos\theta\,q_Z\mathcal Z_+.
\end{aligned}
\end{equation}
Here $A_Z,B_Z$ are the upper--upper opposite-spin amplitudes,
while $\eta_ZX_Z,\eta_ZY_Z$ are their longitudinal
lower-component counterparts. The transverse lower components
are proportional to $\eta_Z(ix_j\pm y_j)$ and carry the same
$A_Z,B_Z$ combinations.

We retain quadratic contributions from the leading Dirac lower
components, using
\begin{equation}\label{eq:small_parameters}
\epsilon_d\equiv\frac{\lambdabarc}{d},
\quad
\epsilon_P\equiv\frac{P}{mc}
=\frac{\lambdabarc P}{\hbar},
\quad
\epsilon\equiv\max\{\epsilon_d,\epsilon_P\}\ll1.
\end{equation}
Since \(\Sigma_i=\operatorname{diag}(\sigma_i,\sigma_i)\) does not
mix upper and lower components of either particle, the
upper--upper and mixed sectors contribute separately to the norm
and spin correlations. The lower--lower sector contributes first
at fourth order in the lower-component amplitudes. Accordingly,
the \(O(\epsilon^4)\) remainders below refer to this sector
expansion within the quadratic-dispersion packet model; they do
not bound corrections to exact positive-energy Dirac propagation.

The transverse integrations require only the normalized Gaussian
moments
\begin{equation}\label{eq:transverse_moments}
\begin{aligned}
\langle f\rangle_\perp
&\equiv\frac{1}{\pi l_Z^2}
\int_{\mathbb R^2}dx\,dy\,
e^{-(x^2+y^2)/l_Z^2}f(x,y),\\
\langle x\rangle_\perp
&=\langle y\rangle_\perp
=\langle xy\rangle_\perp=0,\\
\langle x^2\rangle_\perp
&=\langle y^2\rangle_\perp=\frac{l_Z^2}{2}.
\end{aligned}
\end{equation}
For identical centered circular apertures
\(\lvert\boldsymbol{\rho}_j\rvert<R\), let
\(u=R^2/l_Z^2\). The accepted fraction of the common
   transverse Gaussian is \(1-e^{-u}\) at each plane.
Its conditional radial moment is
\begin{equation}
\bigl\langle \rho^2\bigr\rangle_R
=
l_Z^2\left[1-\frac{u}{e^u-1}\right].
\label{eq:finite_window_moment}
\end{equation}
With \(R=2d\) and the parameters of
Fig.~\ref{fig:prep_sensitive_chsh}, the leading-spin
transverse pair acceptance, conditional on the selected planes,
is approximately \(95.8\%\) at \(Z=d\) and \(74.8\%\) at \(Z=5d\). At leading spin order, the common transverse envelope cancels
from the normalized tensor for these finite apertures.

This cancellation is restricted to leading-spin transverse acceptance.
Longitudinal and timing windows average different assignment amplitudes
and phases; their unnormalized spin numerators and norm must be averaged
before taking the ratio.

The measure factorizes between the two planes.  All common envelope
and scalar propagation factors cancel in
Eq.~\eqref{eq:tau_definition}.  The upper--upper norm is
$|A_Z|^2+|B_Z|^2$; the two mixed sectors together add
$2|\eta_Z|^2[l_Z^2(|A_Z|^2+|B_Z|^2)+|X_Z|^2+|Y_Z|^2]$.
Using
\begin{equation}\label{eq:eta_width_identity}
2|\eta_Z|^2l_Z^2=\frac{\epsilon_d^2}{2},
\end{equation}
the common normalization denominator becomes
\begin{equation}\label{eq:Dfrak_main}
\begin{aligned}
\mathfrak D_Z^{(2)}
&=\left(1+\frac{\epsilon_d^2}{2}\right)
  \left(|A_Z|^2+|B_Z|^2\right)\\
&\quad+2|\eta_Z|^2\left(|X_Z|^2+|Y_Z|^2\right).
\end{aligned}
\end{equation}

Applying the same Gaussian moments to the spin numerators gives
the three independent tensor entries:
\begin{subequations}\label{eq:tau_components}
\begin{align}
\tau_{xx}(Z)
&=\frac{2\Re[-A_Z^*B_Z+2|\eta_Z|^2X_Z^*Y_Z]}
        {\mathfrak D_Z^{(2)}}
  +O(\epsilon^4),
\label{eq:tauxx_main}\\
\tau_{xy}(Z)
&=\frac{2\Im[A_Z^*B_Z-2|\eta_Z|^2X_Z^*Y_Z]}
        {\mathfrak D_Z^{(2)}}
  +O(\epsilon^4),
\label{eq:tauxy_main}\\
\tau_{zz}(Z)
&=-1+\frac{\epsilon_d^2(|A_Z|^2+|B_Z|^2)}
             {\mathfrak D_Z^{(2)}}
  +O(\epsilon^4).
\label{eq:tauzz_main}
\end{align}
\end{subequations}
The transverse lower components produce the correction to the
opposite-spin value $\tau_{zz}=-1$.  Strict expansion of the
normalized ratio gives
\begin{equation}\label{eq:tauzz_expanded_main}
\tau_{zz}(Z)=-1+\epsilon_d^2+O(\epsilon^4),
\end{equation}
so the preparation and separation dependence cancels from this
entry through second order.  
% V3 C03: Keep norm control and state physicality; individual diagnostics and tolerances are in response Point 6.
The normalized expansion requires the leading factored norm
\(S_Z=|A_Z|^2+|B_Z|^2\) to dominate the lower-component correction.
For the three preparations in Fig.~\ref{fig:prep_sensitive_chsh},
\(S_Z>0.539\) and the correction is below \(1.1\times10^{-7}S_Z\)
over \(0\le Z/d\le3\). The state-derived reduced spin matrices are
normalized and positive within numerical roundoff and reproduce the
reported tensor. Thus the plotted suppression and revival are clear
of small-denominator instability.\label{loc:physicality}

The remaining entries follow from the same integration and the
cylindrical symmetry:
\begin{equation}\label{eq:tau_complete_main}
\boldsymbol\tau(Z)=
\begin{pmatrix}
\tau_{xx}(Z)&\tau_{xy}(Z)&0\\
-\tau_{xy}(Z)&\tau_{xx}(Z)&0\\
0&0&\tau_{zz}(Z)
\end{pmatrix}.
\end{equation}
Here $\tau_{xy}=\langle\Sigma_x\otimes\Sigma_y\rangle_Z$ fixes
the off-diagonal sign convention.  For arbitrary unit analyzer
directions $\hat{\mathbf a}$ and $\hat{\mathbf b}$, the correlation is
\begin{equation}\label{eq:C_tensor_main}
\begin{aligned}
C(\hat{\mathbf a},\hat{\mathbf b};Z)
&=\hat{\mathbf a}^{T}\boldsymbol\tau(Z)\hat{\mathbf b}\\
&=\tau_{xx}(a_xb_x+a_yb_y)
 +\tau_{xy}(a_xb_y-a_yb_x)\\
&\quad+\tau_{zz}a_zb_z.
\end{aligned}
\end{equation}

The preparation \((d,P,\theta,\chi)\) and offset \(Z\) therefore
determine the tensor explicitly through
Eqs.~\eqref{eq:qZ_definition}--\eqref{eq:tau_components}.

\section{Fixed and optimized CHSH responses}

For analyzer directions \(\hat{\mathbf a},\hat{\mathbf a}'\) at
detector \(A\) and \(\hat{\mathbf b},\hat{\mathbf b}'\) at detector \(B\), the separation-dependent CHSH combination is
\begin{equation}
\begin{aligned}
B(Z)
={}&C(\hat{\mathbf a},\hat{\mathbf b};Z)
+C(\hat{\mathbf a},\hat{\mathbf b}';Z)\\
&+C(\hat{\mathbf a}',\hat{\mathbf b};Z)
-C(\hat{\mathbf a}',\hat{\mathbf b}';Z)\\
={}&\hat{\mathbf a}^{T}\boldsymbol{\tau}(Z)
       (\hat{\mathbf b}+\hat{\mathbf b}')
+\hat{\mathbf a}'^{T}\boldsymbol{\tau}(Z)
       (\hat{\mathbf b}-\hat{\mathbf b}').
\end{aligned}
\label{eq:CHSH_definition}
\end{equation}
We compare \(B_{\rm fixed}(Z)\) for one four-setting geometry with
\(B_{\max}(Z)\), the maximum CHSH magnitude over local analyzer
directions at each \(Z\). Both use the same conditioned tensor and
\(\boldsymbol\Sigma\)-analyzer model. Their comparison identifies
the separation dependence that survives analyzer reorientation.

\subsection{Prescribed-setting CHSH response}

We hold the following four analyzer directions fixed throughout the
separation scan:
\begin{equation}\label{eq:fixed_analyzer_geometry}
\begin{aligned}
\hat{\mathbf a}&=\hat{\mathbf z},&
\hat{\mathbf a}'&=\hat{\mathbf x},\\
\hat{\mathbf b}
&=\frac{\hat{\mathbf x}+\hat{\mathbf z}}{\sqrt2},&
\hat{\mathbf b}'
&=\frac{-\hat{\mathbf x}+\hat{\mathbf z}}{\sqrt2}.
\end{aligned}
\end{equation}
Substitution into Eqs.~\eqref{eq:C_tensor_main} and
\eqref{eq:CHSH_definition} gives the signed response
\begin{equation}\label{eq:Bfixed_explicit}
B_{\rm fixed}(Z)
=\sqrt2\,[\tau_{zz}(Z)+\tau_{xx}(Z)].
\end{equation}
Thus this geometry samples one specified combination of the
endpoint-conditioned tensor. The optimized response \(B_{\max}(Z)\),
considered below, allows the analyzer directions to vary.

At the coincident-plane boundary \(Z=0\), we have \(A_0=B_0\).
Provided the leading accepted weight is nonzero, the conditioned
leading-spin sector has singlet structure, and hence
\begin{equation}\label{eq:Bfixed_zero}
B_{\rm fixed}(0)=-2\sqrt2+O(\epsilon_d^2).
\end{equation}
This is a limit of the conditioned model, not a Bell test between
separated stations.

At the opposite limit, suppose the endpoint assignments are
momentum resolved:
\begin{equation}\label{eq:separated_limit}
Z\to\infty,\quad
\frac{Pd}{\hbar}\gg1,\quad
q_Z\to\rho_\infty
=\exp[-4(Pd/\hbar)^2]\simeq0.
\end{equation}
The asymptotic expressions below are limits of the specified
quadratic-dispersion model. Their accuracy relative to exact
Dirac evolution must be assessed over finite propagation times;
the small momentum parameters alone do not establish a uniform
error bound as \(Z\to\infty\). The leading-spin result then becomes
\begin{equation}\label{eq:Bfixed_infinity}
B_{\rm fixed}(\infty)
=-\sqrt2\,(1+\gamma_c)
+O(\epsilon^2,\rho_\infty),
\end{equation}
where
\begin{equation}\label{eq:gamma_c}
\gamma_c=\sin(2\theta)\cos\chi
\end{equation}
is the real quadrature of the prepared opposite-spin coherence.

The endpoint values do not describe the intervening scan. At finite
\(Z\), the magnitude and phase of \(q_Z\) change the
endpoint-conditioned tensor and can produce suppression and revival
in \(\lvert B_{\rm fixed}(Z)\rvert\). Whether this variation persists under
optimization is determined from the full tensor in the next
subsection.

\subsection{Optimized-setting CHSH response}

We maximize the CHSH magnitude over analyzer directions at each
\(Z\), using the same endpoint-conditioned tensor and
\(\boldsymbol{\Sigma}\)-analyzer model. In the
upper/lower-sector and spin decomposition of each Dirac space,
\(\Sigma_i=I_{\rm sector}\otimes\sigma_i\). Tracing over the
sector indices and transverse coordinates therefore gives a
normalized two-qubit spin density operator with correlation tensor
\(\boldsymbol{\tau}(Z)\). For
\(C=\hat{\mathbf a}^{T}\boldsymbol{\tau}(Z)\hat{\mathbf b}\),
the maximum is consequently determined by the two largest
eigenvalues of
\(\boldsymbol{\tau}^{T}(Z)\boldsymbol{\tau}(Z)\)
\cite{Horodecki1995}.

At leading spin order, the accepted state is proportional to
\(A_Z|\uparrow\downarrow\rangle-B_Z|\downarrow\uparrow\rangle\).
For \(S_Z=|A_Z|^2+|B_Z|^2>0\), its concurrence and optimized response are
\begin{equation}\label{eq:coherence_CHSH}
\begin{aligned}
C_Z^{(0)}&=\frac{2|A_ZB_Z|}{|A_Z|^2+|B_Z|^2},\\
B_{\max}^{(0)}(Z)&=2\sqrt{1+\bigl[C_Z^{(0)}\bigr]^2}.
\end{aligned}
\end{equation}
Thus direct--exchange interference changes the accepted coherence
through \(A_Z,B_Z\), and this change survives analyzer optimization.
Here \(C_Z^{(0)}\) characterizes the endpoint-conditioned pure spin
state, rather than the source-mode preparation concurrence.

Equation~\eqref{eq:tau_complete_main}
gives two degenerate transverse singular values,
\((\tau_{xx}^2+\tau_{xy}^2)^{1/2}\), and one longitudinal
singular value, \(|\tau_{zz}|\). Hence

{\small
\begin{equation}\label{eq:Bmax}
\begin{aligned}
B_{\max}(Z)=2\Bigl[
&\tau_{xx}^2(Z)+\tau_{xy}^2(Z)\\
&+\max\!\left\{
\tau_{xx}^2(Z)+\tau_{xy}^2(Z),
\tau_{zz}^2(Z)
\right\}
\Bigr]^{1/2}
+O(\epsilon^4).
\end{aligned}
\end{equation}
}

For the quadratic lower-component corrections, using the same \(S_Z\), define
\[
\begin{aligned}
\mathcal T_Z={}&
\frac{4|A_Z^*B_Z|^2}{S_Z^2}
\left[1-\epsilon_d^2
-\frac{4|\eta_Z|^2(|X_Z|^2+|Y_Z|^2)}{S_Z}\right]\\
&-\frac{16|\eta_Z|^2
\Re[(A_Z^*B_Z)^*X_Z^*Y_Z]}{S_Z^2}.
\end{aligned}
\]
Then \(\tau_{xx}^2+\tau_{xy}^2=\mathcal T_Z+O(\epsilon^4)\)
and \(\tau_{zz}^2=1-2\epsilon_d^2+O(\epsilon^4)\), so
Eq.~\eqref{eq:Bmax} becomes
\begin{equation}\label{eq:Bmax_leading_finiteZ}
\begin{aligned}
B_{\max}(Z)
=2\Bigl[\mathcal T_Z
+\max\{\mathcal T_Z,1-2\epsilon_d^2\}\Bigr]^{1/2}
+O(\epsilon^4).
\end{aligned}
\end{equation}
Its zeroth-order transverse singular value is \(C_Z^{(0)}\),
recovering Eq.~\eqref{eq:coherence_CHSH}.

For \(S_Z>0\), the leading pure opposite-spin sector violates CHSH
precisely when \(A_ZB_Z\ne0\). The quadratic expression also retains
the longitudinal and transverse lower-component corrections. The
maximum must remain explicit because the singular-value ordering can
change near maximal coherence; the same nonsuppressed-norm condition
applies.

% V3 C04: Place the compact preparation check beside the leading-spin violation criterion.
For leading-spin point sampling, a numerical search over
\(0\le Z/d\le3\),
\(0.99(\pi/4)\le\theta\le1.01(\pi/4)\), and
\(|\chi-\pi/4|\le\pi/32\) finds a minimum
\(B_{\max}^{(0)}\simeq2.0021\).\label{loc:preparation_tolerance}

At \(Z=0\),
\(A_0=B_0=\cos\theta+e^{i\chi}\sin\theta\). Provided this
amplitude is nonzero and the accepted-norm condition holds,
\(\mathcal T_0=1-\epsilon_d^2\) through second order. Hence
\begin{equation}\label{eq:Bmax_zero}
\begin{aligned}
B_{\max}(0)
&=2\sqrt2\left(1-\frac{\epsilon_d^2}{2}\right)
+O(\epsilon^4),\\
|B_{\rm fixed}(0)|
&=2\sqrt2\left(1-\frac{3\epsilon_d^2}{4}\right)
+O(\epsilon^4).
\end{aligned}
\end{equation}
Both responses reach \(2\sqrt2\) at zeroth order, but their
quadratic corrections differ. These limiting values do not
identify the prepared source spin sector as a singlet. The accepted
upper--upper sector has the singlet structure
\begin{equation}
\mu_0^{(0)}
\propto
\bigl(\cos\theta+e^{i\chi}\sin\theta\bigr)
\bigl(
|\uparrow_A\downarrow_B\rangle
-
|\downarrow_A\uparrow_B\rangle
\bigr).
\end{equation}
The coincident-plane value is a limit of the model, rather than
a Bell test between separated stations.

In the momentum-resolved limit of
Eq.~\eqref{eq:separated_limit}, the residual exchanged
assignment is negligible, so \(A_Z\simeq\cos\theta\) and
\(B_Z\simeq e^{i\chi}\sin\theta\). Since
\(2|\eta_Z|^2|\mathcal Z_-|^2\to\epsilon_P^2/2\),
the transverse singular value squared becomes
\(\sin^2(2\theta)(1-\epsilon_d^2-2\epsilon_P^2)\)
through second order. The condition \(Pd/\hbar\gg1\) implies
\(\epsilon_P\gg\epsilon_d\), so the longitudinal singular
value is the larger one in this limit. The responses are
\begin{align}
B_{\rm fixed}(\infty)
&=-\sqrt2\Bigl[1+\gamma_c
-\left(1+\frac{\gamma_c}{2}\right)\epsilon_d^2
-\gamma_c\epsilon_P^2\Bigr]
\nonumber\\
&\quad+O(\epsilon^4,\rho_\infty),
\nonumber\\
B_{\max}(\infty)
&=2\Bigl[1+\sin^2(2\theta)
-\bigl(2+\sin^2(2\theta)\bigr)\epsilon_d^2
\nonumber\\
&\qquad-2\sin^2(2\theta)\epsilon_P^2\Bigr]^{1/2}
+O(\epsilon^4,\rho_\infty).
\label{eq:Bell_infinity_comparison}
\end{align}
For the illustrated electron parameters \(d=10\,\mathrm{nm}\)
and \(P=5\hbar/d\), \(\epsilon_d\simeq3.86\times10^{-5}\)
and \(\epsilon_P\simeq1.93\times10^{-4}\). For
\(\theta=\pi/4\) and \(\chi=\pi/4\), the retained quadratic
terms reduce the separated-endpoint magnitudes
\(\lvert B_{\rm fixed}\rvert\) and \(B_{\max}\) by approximately
\(4.0\times10^{-8}\) and \(5.6\times10^{-8}\), respectively.
These are lower-component corrections within the packet model,
not error bounds relative to exact Dirac propagation.

Here \(\gamma_c=\sin(2\theta)\cos\chi\), as defined above.
The prescribed geometry samples this real coherence quadrature,
while optimization depends on the coherence magnitude
\(|\sin(2\theta)|\) and the retained lower-component corrections.
For imbalanced preparation, the optimized separated-endpoint
value is below its coincident-plane value in
Eq.~\eqref{eq:Bmax_zero}. In the completely imbalanced case
\(\theta=0\),
\begin{equation}
B_{\max}(\infty)
=2\bigl[1-2\epsilon_d^2\bigr]^{1/2}
+O(\epsilon^4,\rho_\infty).
\end{equation}

Within the retained packet model, the deficit below \(2\)
arises from the Dirac lower-component contributions. For strongly
confined packets, where \(\epsilon_d\) is no longer small, a
quantitative prediction requires a fully relativistic treatment
beyond the scope of this approximation.

\begin{figure}[!tbp]
  \centering
  \includegraphics[width=\columnwidth]{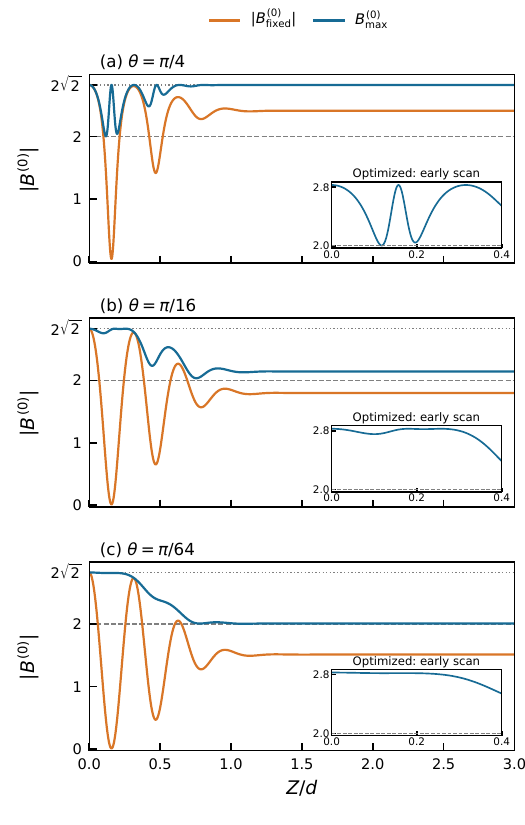}
  \caption{Leading-spin CHSH responses for
  \(d=10\,\mathrm{nm}\), \(P=5\hbar/d\), and \(\chi=\pi/4\):
  (a) \(\theta=\pi/4\), (b) \(\theta=\pi/16\), and
  (c) \(\theta=\pi/64\). Orange:
  \(\lvert B_{\rm fixed}^{(0)}(Z)\rvert\); blue:
  \(B_{\max}^{(0)}(Z)\). Dotted and dashed lines mark \(2\sqrt2\)
  and the CHSH local bound \(2\), respectively. Insets enlarge the optimized response
  over \(0\le Z/d\le0.4\).}
  \label{fig:prep_sensitive_chsh}
\end{figure}

For balanced preparation, the endpoint values coincide at zeroth
order, while their quadratic corrections differ. These endpoint
limits do not imply a flat separation scan. With
\(A_Z=(1+e^{i\chi}q_Z)/\sqrt2\) and
\(B_Z=(e^{i\chi}+q_Z)/\sqrt2\), we have
\begin{align}
|A_Z|^2-|B_Z|^2
&=2\rho_Z\sin\chi\sin\Theta_Z,
\nonumber\\
|A_Z|^2+|B_Z|^2
&=1+\rho_Z^2+2\rho_Z\cos\chi\cos\Theta_Z .
\label{eq:balanced_weights}
\end{align}
The leading squared concurrence is
\[
F_Z=1-
\frac{4\rho_Z^2\sin^2\chi\sin^2\Theta_Z}
{\bigl(1+\rho_Z^2+
2\rho_Z\cos\chi\cos\Theta_Z\bigr)^2}.
\]
Substitution into Eq.~\eqref{eq:Bmax_leading_finiteZ} yields
\begin{equation}\label{eq:Bmax_balanced}
\begin{aligned}
\mathcal T_{Z,\mathrm{bal}}
={}&F_Z\left[1-\epsilon_d^2
-\frac{4|\eta_Z|^2(|X_Z|^2+|Y_Z|^2)}{S_Z}\right]\\
&-\frac{16|\eta_Z|^2
\Re[(A_Z^*B_Z)^*X_Z^*Y_Z]}{S_Z^2},\\
B_{\max,\mathrm{bal}}(Z)
={}&2\Bigl[\mathcal T_{Z,\mathrm{bal}}\\
&+\max\{\mathcal T_{Z,\mathrm{bal}},1-2\epsilon_d^2\}
\Bigr]^{1/2}
+O(\epsilon^4).
\end{aligned}
\end{equation}
Here \(S_Z,A_Z,B_Z,X_Z,Y_Z\) are evaluated at
\(\theta=\pi/4\). Although the prepared branches have equal
weights, direct--exchange interference makes their
endpoint-conditioned upper--upper weights unequal by
\(2\rho_Z\sin\chi\sin\Theta_Z\). This imbalance reduces
\(F_Z\) and produces the leading suppression of \(B_{\max}(Z)\);
analyzer optimization cannot remove it. The response revives
when the imbalance diminishes, with the displayed quadratic
corrections retained. For \(\chi=0\), \(A_Z=B_Z\) and
\(F_Z=1\): the zeroth-order optimized response is flat, while
the lower-component corrections can still vary with separation.

At leading spin order with point sampling and full transverse
acceptance, the normalized responses depend on \(d\), \(P\), and \(Z\)
only through \(Z/d\) and \(Pd/\hbar\). Thus \(d=10\,\mathrm{nm}\) sets
an illustrative physical scale: the point-sampled curves in
Fig.~\ref{fig:prep_sensitive_chsh} are unchanged under
\(d\to sd\), \(P\to P/s\), and \(Z\to sZ\).
The lower-component corrections and experimental length and time
scales still change under this rescaling. 

Figure~\ref{fig:prep_sensitive_chsh} shows the two responses for
\(d=10\,\mathrm{nm}\), \(P=5\hbar/d\), and three preparation
angles. Here \(Pd/\hbar=5\). As \(Z\) changes, the relative
phase \(\Theta_Z\) varies and the exchanged-assignment magnitude
\(\rho_Z\) decreases, producing the interference pattern before
the exchanged contribution becomes negligible.

Both plotted responses show suppression and revival. With
increasing preparation imbalance, the large-\(Z\) optimized response
approaches \(2\) from above, while the prescribed geometry can yield
a magnitude below \(2\). Panel (a) shows that balanced preparation
also has a varying optimized response despite equal leading endpoint values.

% V3 C05: Define finite acceptance once; retain the averaging rule and representative visibility result.
Finite longitudinal and timing acceptance are modeled by uniform
windows \(|z_1-Z|<h\), \(|z_2+Z|<h\), and \(|t-T|<\Delta t\),
with \(T=mZ/P\). Here \(h\) and \(\Delta t\) are spatial and
common-time halfwidths; full transverse acceptance is retained.
The accepted tensor is
\begin{equation}\label{eq:finite_gate_average}
\overline\tau_{ij}(Z)=
\frac{\int d\Lambda\,G_Z(\Lambda)\mathcal M_{ij}(\Lambda)}
     {\int d\Lambda\,G_Z(\Lambda)\mathcal D(\Lambda)}.
\end{equation}
Here \(\Lambda=(z_1,z_2,t)\); \(\mathcal M_{ij}\) and \(\mathcal D\)
are the unnormalized, transverse-integrated spin numerators and norm
density, including envelope weights. The nonnegative \(G_Z\) combines
the spatial windows and normalized common-time distribution, independently
of analyzer settings and outcomes. Optimization follows this averaging.

For balanced preparation, Fig.~\ref{fig:finite_acceptance} shows
a first minimum and revival of approximately \(2.007\) and \(2.354\)
with \(h=0.5\,\mathrm{nm}\), \(\Delta t=5\,\mathrm{fs}\).
Doubling both halfwidths merges this first pair but retains broader
suppression and recovery. Curves start at \(Z/d=0.105\), ensuring
disjoint spatial windows and post-preparation timing for both width
pairs.\label{loc:finite_acceptance}

\begin{figure}[!tbp]
\centering
\includegraphics[width=\columnwidth]{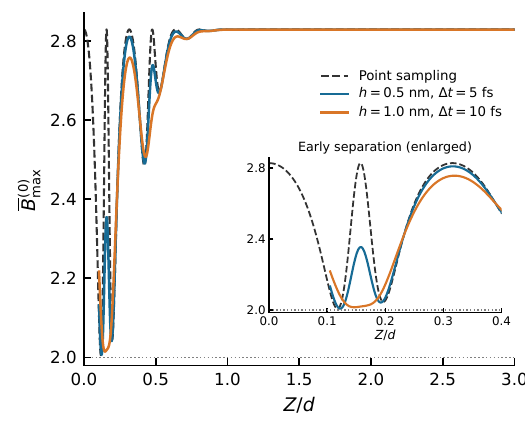}
\caption{Finite-acceptance leading-spin response for the balanced
preparation of Fig.~\ref{fig:prep_sensitive_chsh}(a). Dashed:
point sampling. Solid: uniform spatial and common-time acceptance
with halfwidths \((h,\Delta t)=(0.5\,\mathrm{nm},5\,\mathrm{fs})\)
and \((1\,\mathrm{nm},10\,\mathrm{fs})\), evaluated using
Eq.~\eqref{eq:finite_gate_average}. The inset enlarges
\(0\le Z/d\le0.4\).
Finite-window curves begin at \(Z/d=0.105\), ensuring disjoint spatial
windows and post-preparation timing for both width pairs.
The dotted line marks the CHSH bound \(2\).}
\label{fig:finite_acceptance}
\end{figure}

% V3 C06: Keep one compact matched-parent benchmark; details remain in the response and validation material.
A matched exact positive-energy calculation, with identical initial
Gaussian upper components and lower components matched to leading order,
preserves the balanced preparation's suppression and revival. For the
same sampling and analyzer model, the maximum observed change in
\(B_{\max}\) over \(0\le Z/d\le3\) is approximately
\(9.4\times10^{-9}\).\label{loc:propagation_benchmark}

% V3 C09: End the results section on the physical finding, after its supporting checks.
The separation scan thus resolves how direct--exchange interference
modulates the endpoint-conditioned spin coherence. The resulting
suppression and revival survive analyzer optimization, including for
balanced preparation with equal leading endpoint limits.

% V3 C07: Distinguish conditional spin selection from unitary wave propagation; develop LWR here and retain the quantitative Coulomb boundary.

\section{Discussion}

The separation scan probes one fixed, antisymmetrized
two-Dirac-wave preparation at the endpoints along \(T=mZ/P\).
The spatial extent of the packet branches allows each to contribute
at either station. Along the scan, the direct and exchanged assignments
combine with different relative weights and phases, changing the
sampled opposite-spin amplitudes and the singular values of the spin
tensor. The resulting CHSH suppression and revival survive analyzer
optimization, including for balanced preparation with equal leading
endpoint limits. The complete joint wave evolves unitarily; the
variation concerns its endpoint-conditioned spin state. Prepared
nonseparable coherence and spatial assignment interference are distinct:
the former can persist when the exchanged contribution becomes
negligible.

In WE, this nonseparable joint Dirac wave is a physical entity
carrying the prepared coherence. Spatial separation of the stations
does not turn it into a product of individual electron states.
Free propagation acts separately on the two electron coordinates,
each analyzer is represented by a local \(\boldsymbol\Sigma\)
measurement, and joint outcome probabilities follow the Born rule for
the same setting-independent accepted state. Bell--CHSH supplies the logical benchmark: values
above \(2\) exclude Bell-local outcome factorization when analyzer
choices are independent of the preparation
\cite{Bell1964,CHSH1969,Brunner2014}.
Failure of Bell-local factorization does not by itself establish
transmission between detectors. Here the physical wave parameters
\(d,P,\theta,\chi\) and the sampling geometry determine both the
violation and its interference modulation, without introducing
detector-to-detector transmission. The suppression and revival that
survive optimization provide a quantitative, testable signature of
this coherent Dirac-wave mechanism. Locality here refers specifically
to propagation and local analyzer operations, while reality is assigned to
the joint wave carrying the coherence. This connection provides a
concrete path toward local wave realism (LWR).

An experimental test would hold the preparation fixed while varying
the station separation \(2Z\) and scheduled sampling time \(T=mZ/P\).
The same four prescribed settings determine \(B_{\rm fixed}\), while
reconstruction of \(\boldsymbol\tau\) determines \(B_{\max}\).
Balanced preparation at \(\chi=\pi/4\) tests the optimized modulation,
with the flat leading-spin response \(2\sqrt2\) at \(\chi=0\)
providing a phase control under the same finite acceptance averaging. Accepted pair yields should be recorded alongside correlations
to monitor the changing sampled ensemble.
Figure~\ref{fig:finite_acceptance} quantifies the visibility under
representative finite spatial and timing windows. Comparison with a
particular apparatus requires its actual preparation and spin-sensitive
detector response. The full separation dependence also provides a
quantitative benchmark for future tests of competing physical accounts
under matched preparation, propagation, and sampling conditions.

For \(d=10\,\mathrm{nm}\) and \(P=5\hbar/d\), the short-distance
phase period in \(Z\) is approximately \(3.14\,\mathrm{nm}\), the
central kinetic energy is \(9.5\,\mathrm{meV}\) per electron, and
\(T(Z=d)\simeq173\,\mathrm{fs}\). The balanced example's first minimum
and revival occur near \(20.3\) and \(27.2\,\mathrm{fs}\).
At a characteristic electron separation \(r=d\), the unscreened
Coulomb scale \(e^2/(4\pi\epsilon_0d)\simeq0.144\,\mathrm{eV}\)
exceeds the pair's central kinetic energy \(19.05\,\mathrm{meV}\).
This scale estimate does not justify treating mutual repulsion as
perturbative in the illustrative unscreened realization.
Although spin independent at leading order, the interaction can modify
orbital interference and endpoint-conditioned correlations.
The calculation establishes the free-wavepacket reference and its
specified finite-resolution behavior. An electron implementation must
establish sufficiently weak residual coupling or include the actual
interaction dynamics.\label{loc:coulomb}

\section{Conclusion}

Within the stated free, weakly relativistic packet model, a fixed
Dirac-wavepacket preparation gives separation-dependent Bell
correlations under local endpoint sampling. For the illustrated
\(\chi=\pi/4\) preparations, direct--exchange interference produces
suppression and revival that survive spin-axis optimization.
Even balanced preparation has a varying leading optimized response
between equal endpoint limits of \(2\sqrt2\). Representative finite
spatial and timing windows reduce visibility while retaining broader
modulation. This separation dependence supplies a quantitative,
testable signature of the coherent Dirac-wave mechanism. In WE,
the real, nonseparable joint wave carries the prepared coherence
through deterministic propagation, and local sampling yields the
calculated correlations without postulating transmission between
detectors. The connection between this physical wave account and its
measurable Bell response provides a concrete path toward local wave
realism.

\section*{Conflict of interest}

On behalf of all authors, the corresponding author states that there is no conflict of interest.

\section*{Data availability statement}

No external datasets were used.  The plotted data are generated
directly from the equations given in the manuscript. The numerical results were generated from the equations presented in this article. The validation methods and representative numerical checks are documented in the response to the reviewers.

\bibliographystyle{apsrev4-2}
\bibliography{Bell}

\end{document}